\documentclass[oldversion]{aa} 
\usepackage{graphicx}
\usepackage{txfonts}
\usepackage{natbib}
\bibpunct{(}{)}{;}{a}{}{,}
\begin{document}

   \title{The CORALIE survey for southern extra-solar planets }
   
   \subtitle{XV. Discovery of two eccentric planets orbiting \object{HD~4113} and  \object{HD~156846}.
             \thanks{Based on observations 
             collected with the  {\footnotesize CORALIE} echelle spectrograph on the 1.2-m Euler 
             Swiss telescope at La Silla Observatory, ESO, Chile}}
   
   \author{O.~Tamuz\inst{1,2}   
        \and D. ~S\'egransan\inst{1}    
        \and S.~Udry\inst{1}
        \and M.~Mayor\inst{1} 
        \and A.~Eggenberger\inst{3,1}           
        \and D.~Naef\inst{4} 
        \and F.~Pepe\inst{1}
        \and D.~Queloz\inst{1} 
        \and N.C.~Santos\inst{5} 
        \and B.-O.~Demory\inst{1}   
        \and P.~Figuera\inst{1}  
        \and M.~Marmier\inst{1}  
        \and G.~Montagnier\inst{1,3}  
             }

   \offprints{Damien S\'egransan,     \email{Damien.Segransan@obs.unige.ch}}

   \institute{Observatoire astronomique de l'Universit\'e de Gen\`eve, 
                   51 ch. des Maillettes - Sauverny -, CH-1290 Versoix, 
                   Switzerland
                     \and	     
                   School of Physics and Astronomy, Raymond and Beverly Sackler Faculty of Exact Sciences, Tel Aviv University, Tel Aviv, Israel 
            \and	     
	     Laboratoire d'Astrophysique de Grenoble, BP 53, 38041 Grenoble Cedex
                     \and
             European Southern Observatory, Alonso de Cordova 3107, Casilla
	     19001, Santiago 19, Chile
          \and
             Centro de Astrof\'{\i}sica , Universidade do Porto, Rua das Estrelas, 4150-762 Porto, Portugal 
              }

   \date{Received/accepted}

  \abstract {    We report the detection of two very eccentric planets orbiting  \object{HD~4113}  and \object{HD~156846}  with the 
     {\footnotesize CORALIE} Echelle spectrograph mounted on the 1.2-m Euler Swiss telescope at La Silla. 
    The first planet, HD~4113~b, has minimum mass of $m\sin{i}=1.6\pm0.2~M_{\rm Jup}$, a period of  $P=526.59\pm0.21$~days
    and an eccentricity of $e=0.903\pm0.02$.
    It orbits a metal rich G5V star at $a=1.28$\,AU which displays an additional radial 
    velocity drift of 28~m\,s$^{-1}$/yr observed during 8~years. The combination of the radial-velocity data and the non-detection of any main sequence stellar
    companion in our high contrast images taken at the VLT with NACO/SDI,  characterizes the companion as a probable brown dwarf or as a faint white dwarf.
    The second planet, \object{HD~156846\,b}, has minimum mass of $m\sin{i}=10.45\pm0.05$~M$_{\rm Jup}$, a period of  $P=359.51\pm0.09 $~days,
     an eccentricity of $e=0.847\pm0.002$ and is located at $a=1.0$~AU from its parent star.
    \object{HD\,156846} is a metal rich G0 dwarf and is  also  the primary of a wide binary system ($a>250$\,AU,  $P>4000$\,years). Its stellar companion,   \object{IDS\,17147-1914\,B} , is a M4 dwarf.  The very high eccentricities of both planets can be explained  by Kozai oscillations induced by 
     the  presence of a third object.  
           }


   \maketitle

\section{Introduction}

The  {\footnotesize CORALIE} radial-velocity  planet-search programme  has been on-going for more 
than 9 years (since June 1998) at the 1.2-meter Swiss telescope located at La Silla Observatory, Chile. 
It is a  volume-limited  planet-search survey that contains all main sequence stars from F8 down to K0 within 50 pc 
and has a color-dependant distance limit for later type stars down to M0  \citep{udry-2000:a}. Among the 
1650 stars surveyed, 40 percent of them have a radial-velocity accuracy of 5 m\,s$^{-1}$ or better and 90 percent of the 
sample is monitored with an accuracy better than 10 m\,s$^{-1}$, the limitation being mainly due to photon noise. The 
remaining 10 percent of the sample have a lower accuracy due to the lower signal to noise ratio and/or to the 
fast rotation of the targets.

So far,  {\footnotesize CORALIE} has allowed the detection (or has contributed to the detection) of 48 extra-solar planet candidates 
\citep[e.g.][]{Mayor-2004,Segransan-2007}. 
This substantial contribution together with discoveries from various other programmes have provided a sample of 
more than 230 exoplanets that now permits us to point out interesting statistical constraints for the planet formation 
and evolution scenarios \citep[see e.g.][]{Marcy-2005, Udry-2007:a, Udry-2007:b}, and 
reference therein for reviews on different aspects of the orbital-element distributions or primary star properties).  

 In this paper we report the detection of  two of the  most eccentric known planets ,  \object{HD\,4113\,b} and \object{HD\,156846\,b}.
Together with HD\,80606\,b  \citep{Naef-2001:a} and  HD\,20782\,b \citep{Jones-2006}, only four 
of the known planets have eccentricities larger than 0.8. The possible origin of these eccentricitie is still under debate. 
Suggestions have been made for scenarios which allow the formation of eccentric orbits in a protoplanetary disk 
of planetesimals \citep{Levison-1998} or gas \citep{Goldreich-2003}. 
However, the presence of the third body in both \object{HD\,156846\,b} and \object{HD\,4113\,b}  suggests that eccentricity pumping might
be at work in those two cases \citep{Kozai-1962,Holman-1997,Innanen-1997, Mazeh-1997:a}.
  
The paper is organized as follows. In Sect.\,2 we briefly discuss the primary star properties. Radial-velocity 
measurements and  orbital solutions are presented in Sect.\,\ref{sec:measurements}. In Sect.\,\ref{sec:discussion}, 
we provide some concluding remarks. 
      
\section{Stellar characteristics }
\label{sec:characteristics}   

\subsection{\object{HD\,4113}, (\object{HIP\,3391}, \object{SAO~192693})}   
HD\,4113 is identified as a G5 dwarf in the Hipparcos catalog \citep{ESA-1997} and has an astrometric parallax, $\pi = 22.70\pm 0.99$\,mas 
 which sets the star at a distance of 44.0\,pc from the Sun. With an apparent magnitude $V = 7.88$ 
this implies an absolute magnitude of $M_{V} = 4.66$. According to the Hipparcos catalogue 
the color index for HD\,4113 is $B-V = 0.716$. 
Using a bolometric correction $BC=-0.136$ \citep{Flower-1996} and the solar absolute magnitude M$_{bol}=4.746$ \citep{Lejeune-1998},  we thus obtain a luminosity $L = 1.22$~L$_{\odot}$. 
Stellar parameters for \object{HD\,4113} are summarized in Table~\ref{table1}. 

A detailed spectroscopic analysis of \object{HD\,4113} was 
performed using high signal to  noise  {\footnotesize CORALIE} spectra in order to obtain accurate atmospheric 
parameters \cite[see][for further details]{Santos-2005}.  This gave the following values: an effective temperature 
$T_{\rm eff} = 5688 \pm 26$~K, a surface gravity $\log{g}  = 4.4 \pm 0.05$, and a metallicity $[{\rm Fe/H}] = 0.20 \pm 0.04$. 
Using these parameters and the Geneva stellar evolution models \citep{Meynet-2000}, we 
derive a mass $M = 0.99$\,M$_{\odot}$. According to the Bayesian age estimates of \citet{Pont-2004}, \object{HD\,4113} is an old main-sequence 
star  (4.8-8.0~Gyr).
From the  {\footnotesize CORALIE} spectra 
we derive $v\sin{i} = 1.37$\,km\,s$^{-1}$  \citep{Santos-2002}.

\begin{table}[]
\caption{
\label{table1}
Observed and inferred stellar parameters for the stars hosting planets presented in this paper. Definitions 
and sources of the quoted values are given in the text. 
}
{\center
\begin{tabular}{llcc}
\hline\hline
Parameters                         &                &\object{HD\,4113}   & \object{HD\,156846}       \\
\hline		            			                         
Spectral Type                      &                & G5V                & G0V              \\ 
V                                  &                &7.88                &6.506             \\
$B-V$                              &                &0.716               &  0.557           \\
$\pi$                              & [mas]          & 22.70 $\pm$0.99    &  20.41$\pm$0.93  \\
 $M_{V}$                           &                &4.660               & 3.055            \\
 $T_{\rm eff}$                     & [K]            & 5688$\pm$26        &  6138$\pm$36     \\

$\log{g}$                          & [cgs]          &4.40$\pm$0.05       &4.15$\pm$0.10     \\
$[$Fe/H$]$                         & [dex]          &0.20$\pm$0.04       & 0.22$\pm$0.05    \\
$L$                                & [$L_{\odot}$]  &1.22                &4.98              \\
$M_{\star}$                        & [$M_{\odot}$]  &0.99                &1.43              \\
$v \sin{i}$                      & [km\,s$^{-1}$]         &1.37                & 4.45             \\
Age                                &[Gyr]           &  4.8-8.0           &2.1-2.7           \\

\hline
\end{tabular}
}
\end{table}

 \subsection{\object{HD\,156846} (\object{HIP\,84856} , \object{HR\,6441}, \object{IDS\,17147-1914\,A})}

\object{HD\,156846} is a bright, metal-rich  and  slightly evolved G0 dwarf. The astrometric parallax from the 
Hipparcos catalogue, $\pi=20.41\pm 0.93$\,mas \citep{ESA-1997} sets the star at a distance 
of 49.0 pc from the Sun. With an apparent magnitude $V = 6.50$  this implies 
an absolute magnitude of $M_{V} = 3.055$. According to the Hipparcos catalogue 
the color index for \object{HD\,156846} is $B-V = 0.578$. 
Using the method described in the previous section we 
derive a luminosity $L = 4.98$\,L$_{\odot}$
with a bolometric correction of  $BC=-0.049$. 
Stellar parameters for \object{HD\,156846} 
are summarized in Table~\ref{table1} as well. 

A detailed spectroscopic analysis of \object{HD\,156846} gave the following values: an effective 
temperature $T_{\rm eff} = 6138 \pm 36$\,K, a surface gravity $\log{g} = 4.15 \pm 0.10$, and a metallicity 
$[{\rm Fe/H}] = 0.22 \pm 0.05$. Using these parameters and the Geneva stellar evolution models,  we deduce a mass $M=1.43$\,M$_{\odot}$ . According to the Bayesian age estimates of \citet{Pont-2004}, \object{HD\,156846} is a moderately old main-sequence star (2.1-2.7\,Gyr).
  From the  {\footnotesize CORALIE} spectra we derive $v\sin{i}=4.45$\,km\,s$^{-1}$. 
  
\object{HD\,156846} is the primary star of  a wide binary system \citep[Washington Double Star catalog, hereafter WDS][]{Worley-1996}
with an angular separaion of $\rho=5.1''$ and a position angle PA=75 deg.
Based on its near infrared magnitude \citep[J=9.405, H=8.92, ie.][]{Cabrera-Lavers-2006} and on the stellar
 evolutionnary models of the Lyon group \citep{Baraffe-1998}, its companion,  \object{IDS\,17147-1914\,B} is identified as
 an early M dwarf of mass  $M= 0.59M_{\odot}$.
  Using the positions given in the WDS, we obtain a projected 
binary separation of 250\,AU. This translates into an estimated binary semimajor axis of 315\,AU, using the statistical
relation $a/r=1.261$ \citep{Fischer-1992}.

\section{radial-velocity measurements and orbital solutions}   
\label{sec:measurements}   
   
\subsection{\object{HD\,4113}}

We took 130 spectra of \object{HD4113} from October 1999 to October 2007,
yielding radial-velocity measurements with a typical signal-to-noise ratio of 50 
(per pixel at 550 \,nm) with precision of $\sim 3.5$\,m\,s$^{-1}$. 

\begin{figure}[th!]
      \includegraphics[angle=0,width=0.5\textwidth,origin=br]{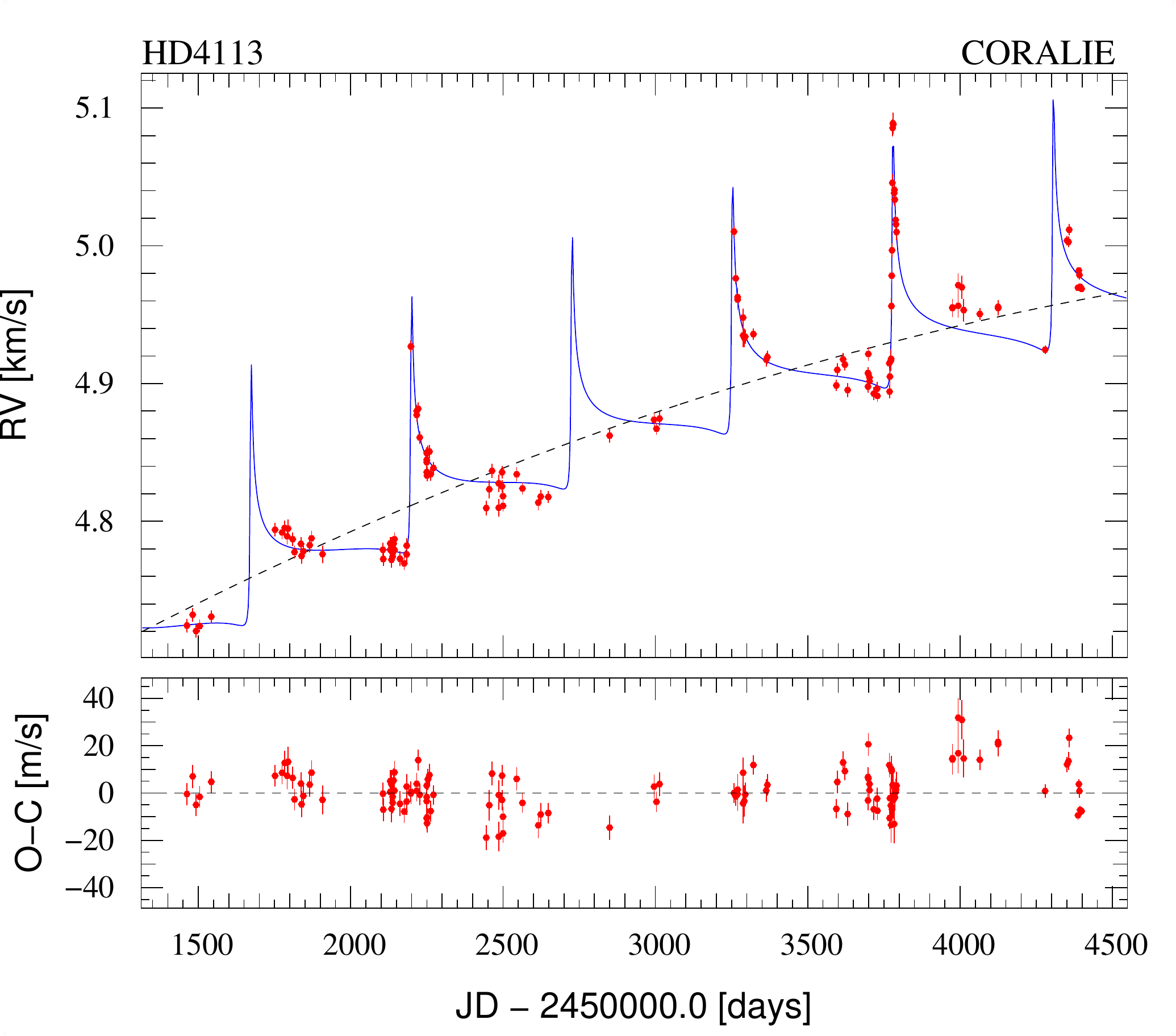}
      \includegraphics[angle=0,width=0.5\textwidth,origin=br]{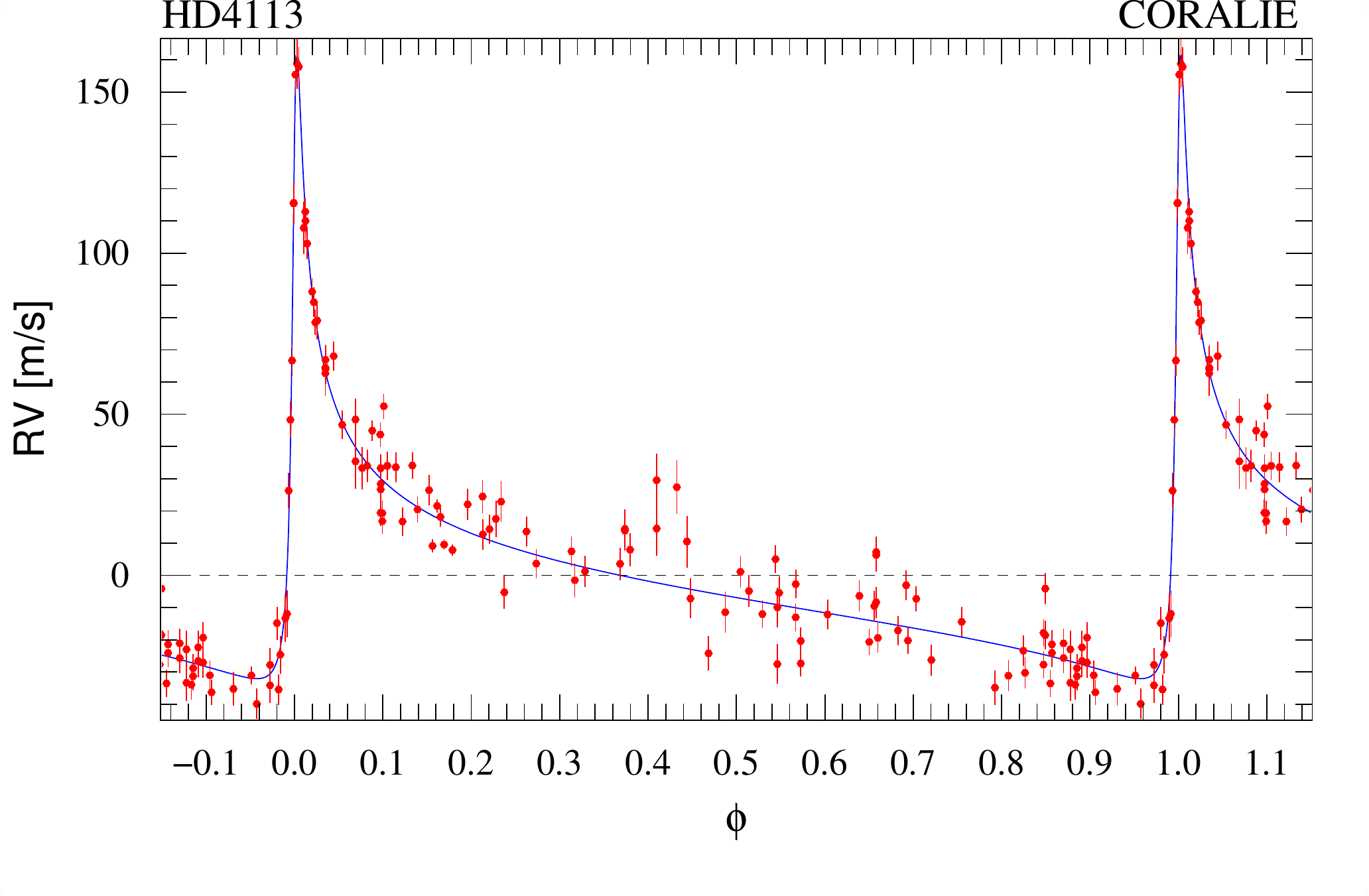} 
\caption[]{
\label{fig:HD4113}
Radial-velocity measurements  as a function of Julian Date obtained with {\footnotesize
CORALIE} for {\footnotesize \object{HD~4113}} superimposed on the best Keplerian planetary solution 
(top figure). The residuals are displayed at the bottom of the top figure and the phase folded radial-velocity 
measurements are displayed on the bottom diagram.}
\end{figure}

Figure~\ref{fig:HD4113} shows the  {\footnotesize CORALIE} radial velocities  and the  corresponding best-fit Keplerian 
model with a linear drift. The resulting orbital parameters are $P = 526.58$\,days, $e = 0.90$, 
$K = 97.7$\,m\,s$^{-1}$, imply the presence of a planet of  minimum mass $m\sin{i}=1.63$\,M$_{\rm Jup}$ orbiting with a 
semi-major axis $a =1.28$\,AU. The orbital separation ranges from 0.12\,AU at periastron to 2.4\,AU at apoastron. 
Residuals are somewhat larger than the precision of the measurements ($\sigma_{(O-C)}=8.4 $\,m\,s$^{-1}$), and are probably  caused by a combination of stellar jitter and the inadequate modeling of the long period companion.

 In addition to the Keplerian signal, the radial-velocity measurements 
display a 28\,m s$^{-1}$ yr$^{-1}$ linear drift with a curvature of over -1.55\,m s$^{-1}$ yr$^{-2}$ over a time span of 8 years induced by a companion at a larger separation.
Based on the duration and the amplitude of the drift, we are able to make some estimates on the nature of the outer companion of \object{HD~4113}.
The lowest possible companion mass responsible for the drift has  $m\sin{i}=10$\,M$_{\rm Jup}$ which correspond to an 
orbital period of 11.5 year, an eccentricity of $e=0.5$ and a radial-velocity semi-amplitude of 150\,m\,s$^{-1}$. However, this orbital solution is 
very unlikely since the outer planet-star separation would reach 2.5\,AU at periastron which is the separation of the
 inner planet at apoastron. More realistic solutions therefore have  longer periods. For instance, the shortest orbital period for a circular 
 orbit is 22 years and correspond to a probable brown dwarf companion ( $m\sin{i}=14$\,M$_{\rm Jup}$) orbiting \object{HD~4113} at  $a=7.9\,AU$. The companion would 
 reach the hydrogen burning limit at a  separation of 20\,AU with a period of 90 years. Radial velocities allow to conclude that the companion responsible for the observed drift is unlikely to be a planet (orbit stability reasons) but cannot discriminate between a brown dwarf and a stellar companion.\\

To bring additional constraints on the outer companion's nature
  we obtained high contrast  images of the target using the VLT and its adaptive optics system NACO in  differential mode \citep[][]{Montagnier-2007}.
We did not detect any object within a radius of 2.5 arcsec,  which excludes any main sequence stellar companions to \object{HD~4113} down to 0.2 arcsec ( about 8\,AU), (Montagnier et al.,  in prep.).  We conclude that the companion responsible for the observed drift is likely to be either a faint white dwarf like  GJ~86~B    \citep[][]{Mugrauer-2005, Lagrange-2006} that went undetected in our SDI images or it could be a brown dwarf located between 8 and 20\,AU from the parent star with a period ranging from 20-90 years.

\subsection{\object{HD\,156846}}

\begin{figure}[th!]
             \includegraphics[angle=0,width=0.5\textwidth,origin=br]{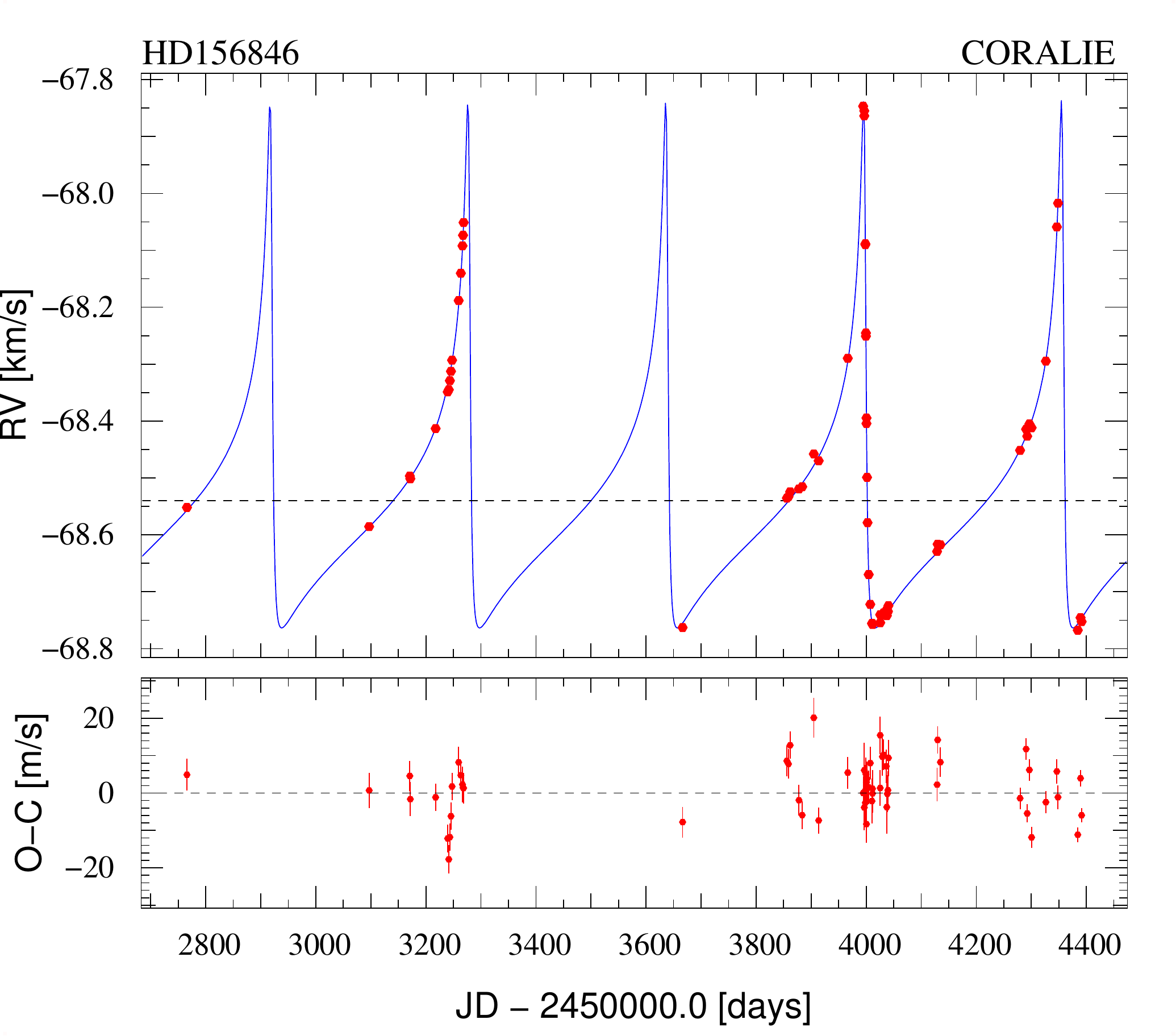}
              \includegraphics[angle=0,width=0.5\textwidth,origin=br]{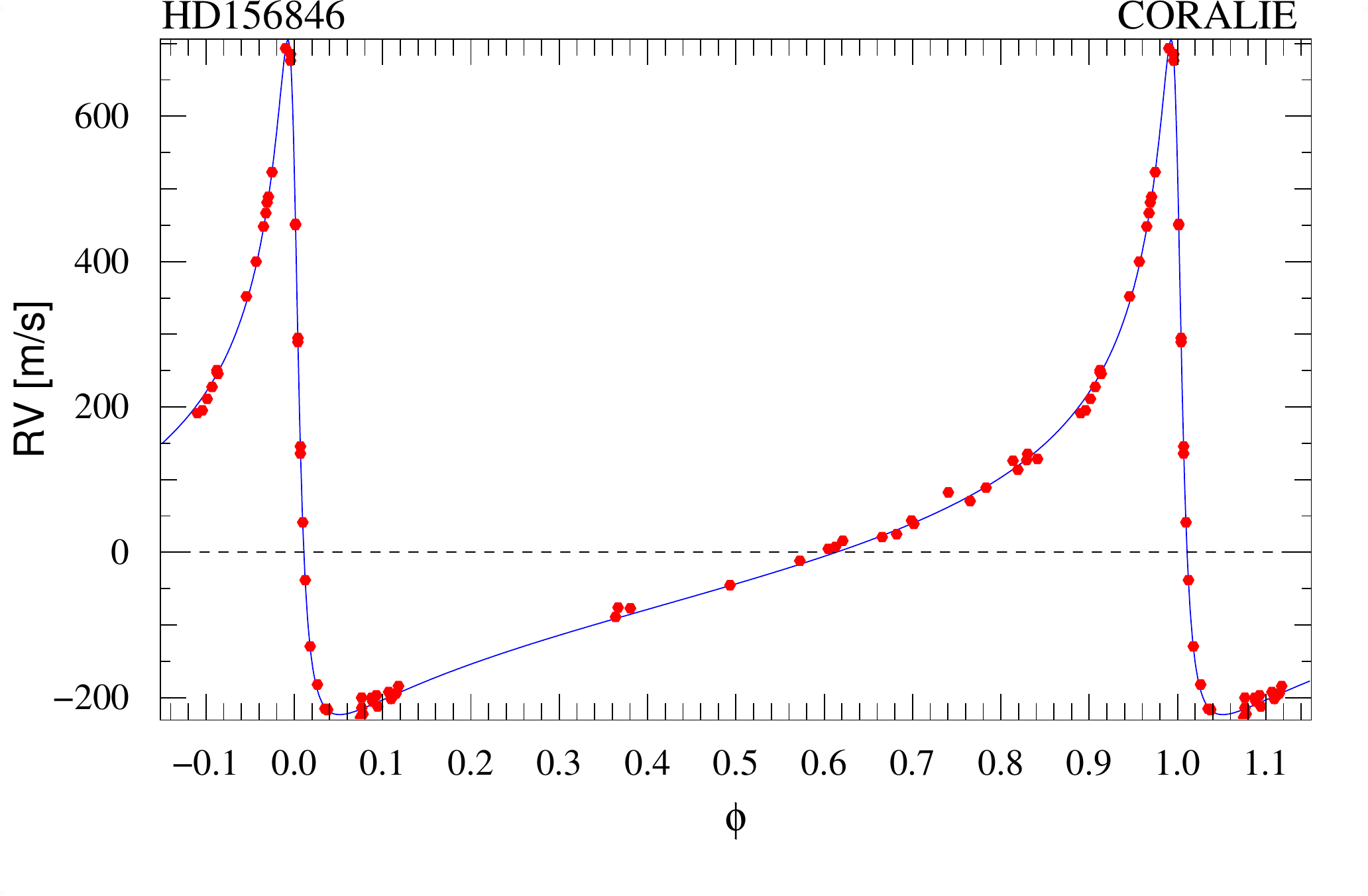}  
 \caption[]{
\label{fig:HD156846}
Same as Fig.~\ref{fig:HD4113} for  {\footnotesize \object{HD\,156846}}.}
\end{figure}

\object{HD\,156846}  has been observed with  {\footnotesize CORALIE} at La Silla Obervatory since May 2003. Altogether, 
64 radial-velocity measurements with a typical signal-to-noise ratio of 75 (per pixel at 550\,nm) and a 
mean measurement uncertainty (including photon noise and calibration errors) of 2.8\,m\,s$^{-1}$
were gathered. Figure~\ref{fig:HD156846} shows the  {\footnotesize CORALIE} radial velocities  and the  corresponding 
best-fit Keplerian model. The resulting orbital parameters are $P = 359.50$\,days, $e = 0.847$, 
$K = 464$\,m\,s$^{-1}$ , implying a minimum mass $m_{2}\sin{i}=10.45$\,M$_{\rm Jup}$ orbiting \object{HD\,156846} 
with a semimajor axis $a =0.99$\,AU. The orbital separation ranges from 0.15\,AU at periastron to 1.8\,AU at apoastron. 
The orbital elements for \object{HD\,156846\,b} are listed in Table~\ref{table:elements}. 
Residuals around the single planet orbital solution are  larger than the precision of the measurements ($\sigma_{(O-C)}=7.5$\,m\,s$^{-1}$) which points
toward the  stellar jitter since the the high eccentricity of \object{HD~156846~b}
does not leave much room for a short period undetected planet.

\begin{table}[]
\caption{
\label{table:elements}
{\footnotesize CORALIE} best Keplerian orbital solutions for \object{HD~156846} and \object{HD~4113}, 
as well as inferred planetary parameters. Confidence intervals are computed for a  68\% confidence 
level  after 10000 montecarlo iterations. {\it Span} is the time interval in days between the first and last
measurements. $\sigma(O-C)$ is the weighted r.m.s. of the residuals around the derived solutions}
\label{table:fit}
\begin{center}
\begin{tabular}{llccc}
\hline\hline
Parameters           &                              &\object{HD~4113}           & \object{HD~156846}                 \\
\hline						                                
\\[-2mm]					                                
$\gamma$             & [km  s$^{-1}$]               &$4.874\pm0.05$          &$-68.540\pm0.001$      \\[1mm]
slope                & [m s$^{-1}$ yr$^{-1}$]       &        $27.8\pm0.3$        & -                      \\[1mm]
curvature            & [m s$^{-1}$ yr$^{-2}$]       &        $-1.55\pm0.17$      & -                      \\[1mm] 
\hline						                                
\\[-2mm]					                                
$P$                  & [days]                       & $   526.62\pm0.3    $     &$359.51\pm0.09$        \\[1mm]
$K$                  & [m\,s$^{-1}$]                &    $  97.1\pm3.8$          &$464.3\pm3.0$          \\[1mm]
$e$                  &                              &  $    0.903\pm0.005$       &$0.8472\pm0.0016$      \\[1mm]
 $\omega$            & [deg]                        &  $     -42.3\pm1.9 $       &$52.23\pm0.41$         \\[1mm]
$T_{0}$              &[JD-2.45$\,10^{6}$]           & $ 3778.0\pm0.2$          &$3998.09\pm0.05$       \\[1mm]
 \hline						                                
$a_{1}\, \sin{i} $      & [$10^{-3}\, $AU    ]      &       $2.01\pm0.05$                 &        $8.15\pm0.04$           \\
$f(m)$               & [$10^{-9}\,$ M$_{\odot}$  ]  &       $3.9\pm0.3$                 &      $558\pm9 $            \\
 $m_{2}\, \sin{i}$   & [M$_{\rm Jup}$]                  &     $1.56\pm0.04$                   &$10.45\pm0.05$                  \\
  $a$                & [AU    ]                   &          1.28             &          0.9930         \\
						                                
\hline  					                                
$N_{mes}$            &                              & 130                        & 64                    \\
{\it Span}           &[years]                       &8.0                         &4.45                    \\
$\sigma_{(O-C)}$     & [m\,s$^{-1}$]                        & 8.4                        &  7.7                 \\
\hline
\end{tabular}
\end{center}
\end{table}

\section{Discussion \& conclusion}
\label{sec:discussion}   
\begin{figure}[!h]
\label{fig:e_vs_logp}
\center
\includegraphics[width=0.8\columnwidth]{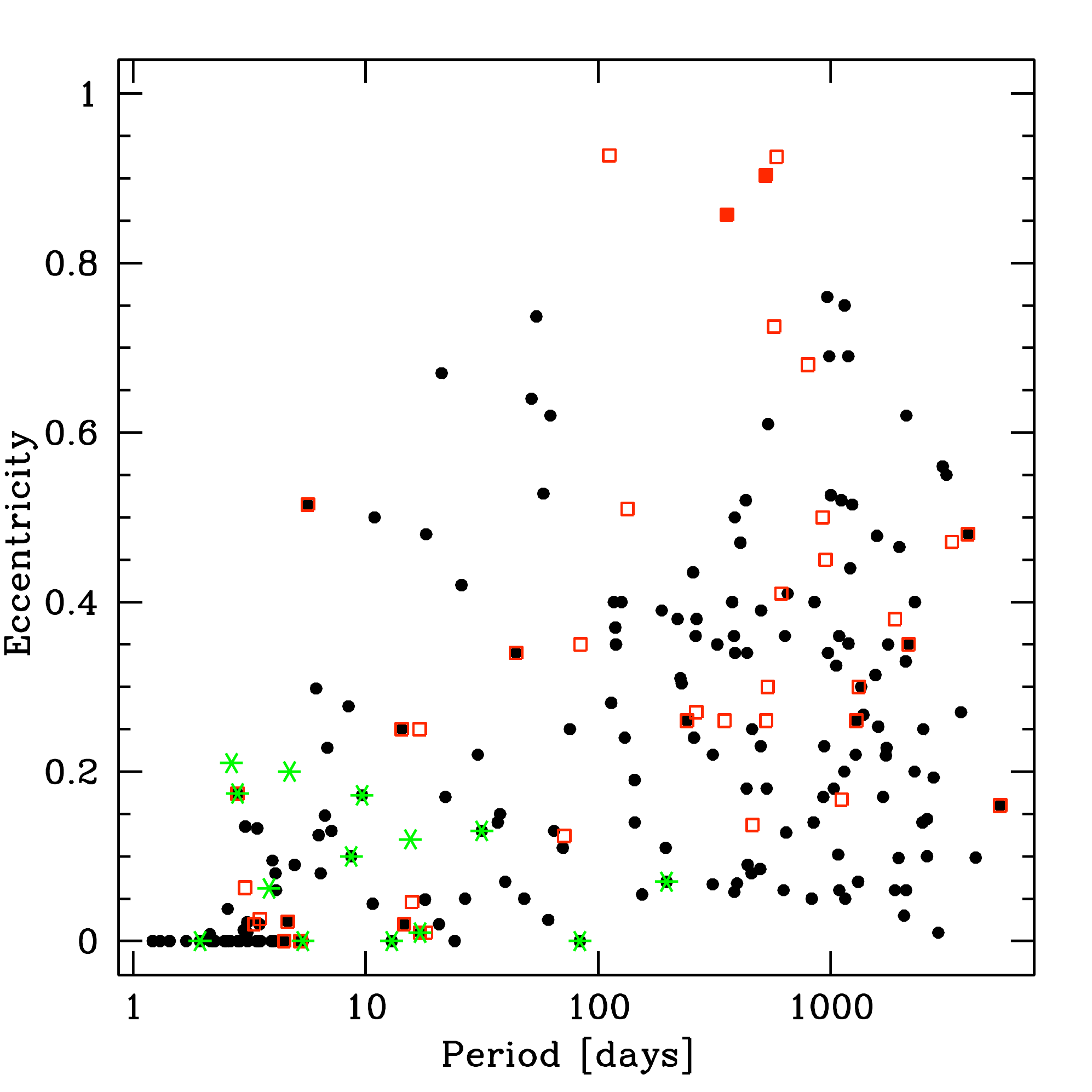}
 \caption{$e$ vs. $\log{P}$ for the known extra-solar giant planets orbiting dwarf primaries. Filled dots are for  planets orbiting single stars and open squares for planets orbiting a component of a multiple stellar system. Green stars correspond to stars hosting Netpune-like planets.
 \object{HD\,4113\,b} and \object{HD\,156846\,b} are located by filled red squares.}
 \end{figure}

 In this paper, we have reported the detection of two very eccentric planets 
  orbiting two metal rich G dwarfs - \object{HD~4113} and \object{HD~156846}  - for which a third massive body is present in each planetary system. 
   The origin of the wide range of eccentricities in extrasolar planets (represented in  Fig.~\ref{fig:e_vs_logp} as a $e$ vs. $\log{P}$ diagram)
  is still under debate, and different scenario have been proposed to explain the largest eccentricities such as the Kozai oscillations \citep{Kozai-1962}, chaotic evolution of planetary orbits in multiple systems and formation of eccentric orbits in a protoplanetary disk 
of planetesimals \citep{Levison-1998} or gas  \citep{Goldreich-2003}. 
  
   \object{HD~156846\,b} and \object{HD~4113\,b} are two of the four most eccentric known planets, together with HD\,80606\,b 
\citep{Naef-2001:a} and HD\,20782\,b  \citep{Jones-2006}  with eccentricities larger than 0.85.
It should be noticed that the parent stars of these 4 planets are part of wide binary systems \citep[][for HD\, 20782 and respective discovery papers for  HD80606, \object{HD~4113} and \object{HD~156846}]{Desidera-2007} raising the question 
of the influence of the third body  on the planets orbital parameters.   It is therefore tempting to investigate  the eccentricity 
 pumping scenario a bit further by computing the different time-scales involved.

 The Kozai mechanism is effective at very long range, but its
oscillations may be suppressed by other competing sources of
orbital perturbations, such as general relativity effects or perturbations
resulting from the presence of an additional companion
in the system. Regarding \object{HD~4113} and \object{HD~156846} , we have estimated
 their Kozai oscillation periods using Eqs (36)  of \citet{Ford-2000} .
 This yields to $P_{Kozai} =  5.5\, 10^6$ years  for \object{HD~156846} and
   $P_{Kozai} =  6.2\, 10^3$ years for \object{HD~4113} which indicates
   that both planet eccentricity can undergo Kozai oscillations with periods 
   that are many order of magnitude shorter than the age of the systems.
   We also computed  the apsidal precession period due to relativistic corrections to the newtonian equation of motion
   using  Eq. (4) of \citet{Holman-1997}.   This yields $P_{GR} =  1.1 $Gyr and$P_{GR} =  7.6$Gyr  for \object{HD~156846} and
   for \object{HD~4113} respectively. This is also several order of magnitude larger that the Kozai oscillation period.
   
   Provided that the orbital planes of the inner planet and the outer companion have different inclinations, the eccentricities of both 
   \object{HD4113~b} and \object{HD~156846~b}  could undergo  Kozai oscillations. That would be the most likely explanation for such high eccentricities.
  
\begin{acknowledgements}

 We are grateful to the staff from the Geneva
Observatory, in particular to L.Weber, for maintaining the 1.2-m Euler
Swiss telescope and the CORALIE echelle spectrograph at La Silla,
and for technical support during observations.
 We thank the Swiss National Research
Foundation (FNRS) and the Geneva University for their continuous
support to our planet search programmes.
We would like to thank T. Mazeh for usefull discussions and the referee for his thoughtful comments.
NCS would like to thank the support from Funda\c{c}\~ao
para a Ci\^encia e a Tecnologia, Portugal, in the form of a
grant (reference POCI/CTE-AST/56453/2004), as well as
support by the EC's FP6 and by FCT (with POCI2010 and
FEDER funds), within the HELAS international collaboration.
Support from the Funda‹o para Cincia e a Tecnologia (Portugal) to P. F. in the form of a scholarship (reference SFRH/BD/21502/2005) is gratefully acknowledged.
  A.E.  acknowledges support from the Swiss National Science Foundation through
a fellowship for prospective researcher.
 This research has made use of the VizieR catalogue access tool operated at
CDS, France.
\end{acknowledgements}

\bibliographystyle{aa}
\bibliography{biblioplanets.bib}

\end{document}